\documentclass[12pt]{article}
\usepackage{amsmath,amssymb,graphicx,mathrsfs,hyperref,slashed}
\newcommand{\be}{\begin{equation}}
\newcommand{\ee}{\end{equation}}
\newcommand{\bea}{\begin{eqnarray}}
\newcommand{\eea}{\end{eqnarray}}

\newcommand{\wt}{\widetilde}

\newcommand{\dn}{\Delta N}
\def\({\left(} \def\){\right)}

\renewcommand{\baselinestretch}{1.15}
\begin{document}
%%%%%%%%%%%%%%%%%%%%%%%%%%%%%%%%%%%%%%%%%
\title{\vspace{-1.8in}
%\begin{flushright} {\footnotesize CERN-PH-TH/2011-204}  \end{flushright}
%\vspace{3mm}
\vspace{0.3cm} {Firewalls, smoke and mirrors}}
\author{\large Ram Brustein${}^{(1)}$,  A.J.M. Medved${}^{(2)}$ \\
 \hspace{-1.5in} \vbox{
 \begin{flushleft}
  $^{\textrm{\normalsize
(1)\ Department of Physics, Ben-Gurion University,
    Beer-Sheva 84105, Israel}}$
$^{\textrm{\normalsize (2)  Department of Physics \& Electronics, Rhodes University,
  Grahamstown 6140, South Africa }}$
 \\ \small \hspace{1.7in}
    ramyb@bgu.ac.il,\  j.medved@ru.ac.za
\end{flushleft}
}}
\date{}
\maketitle
%%%

 \begin{abstract}
The radiation emitted by a black hole (BH) during its evaporation has to have some degree of quantum coherence to accommodate a unitary time evolution. We parametrize the degree of coherence by the number of coherently emitted particles $N_{coh}$ and show that it is severely constrained by the equivalence principle.  We discuss, in this context, the fate of a shell of matter that falls into a Schwarzschild BH.  Two points of view are considered, that of a stationary external observer and  that of the shell itself. From the perspective of the shell, the near-horizon region has an energy density proportional to $N_{coh}^2$ in Schwarzschild units. So, if $N_{coh}$ is parameterically larger than the square root of the BH entropy
$S_{BH}^ {1/2}$, a firewall or more generally a ``wall of smoke" forms and the equivalence principle is violated while the BH is still semiclassical. To have a degree of coherence that is parametrically smaller than $S_{BH}^{1/2}$, one has to introduce a new sub-Planckian gravitational length scale, which likely also violates the equivalence principle. And so our previously proposed model which has $N_{coh}=S_{BH}^{1/2}$ is singled out. From the external-observer perspective, we find that the time it takes for the information about the state of the shell to get re-emitted from the BH is inversely proportional to $N_{coh}$. When the rate of information release becomes order unity, the semiclassical approximation starts to break down and the BH becomes a perfect reflecting information mirror.
\end{abstract}
%\maketitle***
\newpage
\renewcommand{\baselinestretch}{1.5}\normalsize
%\section{Introduction}

The black hole (BH) information-loss paradox  underscores the apparent violation of unitarity in the process of BH evaporation \cite{Hawk,info} (for reviews, see \cite{info2,info3,info4}). This paradox  has recently been recast in a way that flips the logic around. In this modern take, one assumes that the information does come out in the end and then asks as to the consequences for an in-falling observer. It has been argued that such an observer encounters a sea of high-energy quanta, which has been metaphorically dubbed the ``firewall'' \cite{AMPS} (also see \cite{Sunny1,Braun,Mathur}). It is fair to say that this assertion has caught the attention of  the high-energy community
({\em e.g.}, \cite{fw1,fw2,fw3,fw4,Sunny2,avery,lowe,vv,pap,AMPSS,lowefw2,SMfw,pagefw,VRfw,MP,bousso2,newfw1,
newfw2,newfw3}).

The ensuing discussions revolve mostly around the various assumptions adopted by the original firewall proponents
\cite{AMPS}, which are based on the tenets of BH complementarity as expressed in \cite{sussplus}. It is our view, however, that  the critical assumption is a hidden one which we ascribe to the older discussion by Page \cite{page}.  Page assumes that the  density matrix of the BH and the radiation  represents a pure state in some random basis. Then a random unitary transformation
relates this basis  to that in which the density matrix has a single entry. The entries of the matrix are therefore  distributed according to Levy's lemma.  A consequence of this assumption is that the first bit of information emerges at the so-called ``Page time''  when the BH has lost half of its original entropy to the radiation. Once this assumption is adopted in addition to the original assumptions in \cite{AMPS}, a firewall is indeed an inevitable consequence.

To describe semiclassical gravitational physics in a consistent way, one needs to preserve  the equivalence principle in addition to unitarity. It is in this sense that the information-loss paradox and the firewall problem can be seen as two sides of the same problem: Combining general relativity with quantum mechanics while preserving the basic principles of both.

The equivalence principle can be probed by  considering the fate of an in-falling shell onto an existing BH. The fate of the shell will be seen as different depending on what observer is concerned. This is true simply on the basis of classical general relativity  and can be attributed to the effect of boosting between the different reference frames. The two extreme vantage points are that of an external, stationary observer who is stationed far away from the horizon and an observer who is in the rest frame of the freely-falling shell.  The equivalence principle asserts that they have to agree on the results if they use a small enough region in spacetime and  their findings can be compared. A consequence of the equivalence principle is therefore that any differences in the results of the two observers are controlled by the Schwarzschild radius $R_S$ (or the Hawking temperature $T_H$), the only scale in the problem. In the original discussion by Hawking, the equivalence principle is automatically preserved and unitarity is abandoned. In the Page description, unitarity is taken as a given; however,  the appearance of a firewall signals a violation of the equivalence principle \cite{bousso2}.

Given a unitary process of BH evaporation, the fate of the in-falling   shell is quite clear as far as an external observer is concerned. The shell  will thermalize at the horizon and eventually the information about its state will turn up --- albeit, in a highly scrambled form --- in the Hawking radiation. The pertinent question is then  how long must the observer wait to retrieve information about the state of the shell?

Hayden and Preskill (HP) answered this question \cite{HaydenPreskill} for the Page model.
They considered dropping a $k$-bit quantum object into a BH. Quantum in this context  means a finite chain of ordered bits, so that the object is necessarily small in terms of both spatial extent and mass. We will  later consider a more general thought experiment that allows for a macroscopic shell to be dropped. A further reasonable assumption made by HP is that the object thermalizes at the horizon within the order of one scrambling time, $\;t_{scram}\sim R_S\ln{R_S}\;$.  According to HP, the answer depends on the age of the BH that is absorbing the shell. If the BH is old --- meaning after the Page time --- then  according to their quantum-information analysis, a sufficiently  skilled observer can ascertain the state of the object to within an error of $2^{-c}$ by  reading the next $k+c$ bits of information that are emitted by the BH. If the BH is young --- meaning before the Page time --- the observer has to wait until the Page time and  will then be able to reconstruct the state of the shell almost immediately.   That is, a  Schwarzschild BH can  serve as a perfect mirror for  quantum objects.

Recently, we have proposed a  novel model of BH evaporation \cite{slowleak,slowburn,flameoff} that improves
upon the  standard model of Hawking \cite{Hawk,info} in three aspects: (i) The quantum fluctuations of the collapsing shell (or BH) spacetime are included. We do this, following ideas from \cite{RB}  (also see \cite{RM,flucyou} as
well as \cite{RJ,Dvali1,Dvali2,Dvali3}), by endowing the collapsing shell  with a wavefunction and calculating expectation values where appropriate. Only the standard rules of quantum mechanics are applied. A new parameter appears, the width of the wavefunction, which encodes the strength of the quantum corrections.
It is proportional to $\hbar$  and turns out to be equal to $\;C_{BH}=1/S_{BH}=\hbar G_N/ (\pi R_S^2)\;$ in Schwarzschild units. We also take into account
(ii) the time dependence of the emission process  and (iii) the back-reaction on the shell by the emitted particles. These steps are implemented by assigning the wavefunction of the shell a time dependence that is deduced from the thermal rate of particle emissions and the corresponding shrinking of the Schwarzschild radius.

A surprising element that emerges  from our analysis is a new time scale, the coherence time $t_{coh}$. This scale determines, at any given time, the number of coherently emitted radiation modes \cite{slowburn} and the number of entangled Hawking pairs \cite{flameoff}. The number  was called $N_{coh}$ and, in our model, turns out to scale as  $\;N_{coh}\sim \sqrt{S_{BH}}\;$.  It was revealed in \cite{slowburn}  that nearly all of the BH information is released  in the final interval of $t_{coh}$  before the end of evaporation. We called the beginning of
this interval the ``transparency time''.  It  was  established in \cite{flameoff} that the negative-energy interior modes and their exterior partners maintain nearly maximal entanglement up until this moment.

We have found a density matrix for the radiation with many zeros (actually, exponentially suppressed matrix elements). This property  can be contrasted with the ``typical" density matrix that appears in the Page model of BH evaporation, as discussed
above. Without any physical input, the proliferation of zeros would appear to require  ``fine tuning". However, in our model, this specific form of  density matrix can be attributed to the presence of a   horizon.
For a classical geometry, correlations across the horizon will decay to an exponentially small magnitude in an exponentially short time. This is the physical reason that the Hawking model leads to a density matrix that has zeros everywhere except on the diagonal. This
is almost true semiclassically; although some horizon-crossing  correlations decay as a power law, most still decay exponentially and this is the physical origin of the many zeros in our density matrix.

Let us now consider the  fate of an additional shell of matter that falls into the BH. We would first like to consider
 the shell's own recount of
its destiny. In actuality, what we (and all others) mean by this is that  quantities are calculated from a stationary, external observer's point of view and then transformed, using the equivalence principle, to a frame of reference
that is  co-moving with the shell. When doing so, one relies on  the equivalence between an accelerated observer and an observer in a gravitational field. Hence, we will be estimating  what a stationary observer thinks that the shell should see. To determine what the shell really sees after it crosses the horizon would require a model of the BH interior.

We need to calculate the energy density  that the shell sees as it approaches the horizon, as just discussed. The wavefunction of the produced Hawking pairs was calculated in \cite{flameoff}.  The multi-particle (MP) state $\Psi_{SC}^{MP}(N_T;N',N'')$, when  expressed in  Dirac  notation, is given by
\bea
&& |\Psi_{SC}^{MP}(N_T;N',\omega_{out};N'',\wt{\omega}_{in})\rangle = \frac{1}{\sqrt{N_{coh}}}\Biggl[
\frac{1}{\sqrt{e^{\frac{\hbar \omega}{T_{H}}}-1}} \delta_{N',N''} \delta (\omega_{out}-\wt{\omega}_{in})|N'\rangle | N''\rangle \hspace{.5in}\cr
&+&
C^{1/2}_{BH}(N_T)\Delta\rho_{OD}(\omega_{out},\wt{\omega}_{in})\;D(N_T;N',N'')
\left[1-\delta_{N',N''}\delta (\omega_{out}-\wt{\omega}_{in})\right]|N'\rangle| N''\rangle \Biggr]\;,
\label{nmatrix3}
\eea
where $N_T$ is the ``time'' coordinate in terms of the total number of so-far produced pairs
 and $N'$,$N''$ label the pair-production times.
The off-diagonal factors $D(N_T;N',N'')$  are effectively equal to one when either $\;N_T-N_{coh}\lesssim N^{\prime} \leq N_T\;$ or $\;N_T-N_{coh}\lesssim N^{\prime\prime} \leq N_T\;$ and are otherwise vanishing (exponentially suppressed). The normalization is  fixed by  requiring that the reduced density matrix
for the out-modes $\;\rho_{out}=Tr_{in} |\Psi_{SC}^{MP}\rangle\langle\Psi_{SC}^{MP}|\;$  be correctly normalized.

The off-diagonal matrix $\Delta\rho_{OD}(\omega_{out},\wt{\omega}_{in})$ was calculated in \cite{flameoff} and can be treated as a uniform matrix within the thermal window, when $\;\omega_{out}$, $\wt{\omega}_{in}\sim T_H$. The off-diagonal coefficients in $|\Psi_{SC}^{MP}(N_T;N',\omega_{out},N'',\wt{\omega}_{in})\rangle $ behave in a somewhat surprising way. When a Hawking mode is created, it has some small degree of coherence, proportional to $\sqrt{C_{BH}}$,  with all other particles that were  previously emitted. But, after the passing of $t_{coh}$  from the mode's creation, these correlations decay exponentially fast and essentially vanish.

To find the energy density near the horizon,  it is necessary to evaluate the expectation value of the Hamiltonian. The Hamiltonian  is $\;H=i\hbar \partial/\partial u\;$, where $u$ is the retarded time coordinate, $\;u=t-r-2M \ln (r-2M)\;$.  The action of the Hamiltonian on the wavefunction is simple,
\be
H|\Psi_{SC}^{MP}(N_T;N',\omega_{out},N'',\wt{\omega}_{in})\rangle = \left(\omega_{out}-\wt{\omega}_{in}\right) |\Psi_{SC}^{MP}(N_T;N',\omega_{out},N'',\wt{\omega}_{in})\rangle\;.
\ee

This allows us to estimate the energy density per pair ${\cal E}_{\rm pair}$,
 \bea
 {\cal E}_{\rm pair}&=&  \langle \Psi_{pair}(\omega_{out},-\wt{\omega}_{in})| H |\Psi_{pair}(\omega_{out},-\wt{\omega}_{in})\rangle
 \cr &=& \langle \Psi_{pair}(\omega_{out},-\wt{\omega}_{in})| \left(\omega_{out}-\wt{\omega}_{in}\right) |\Psi_{pair}(\omega_{out},-\wt{\omega}_{in})\rangle\;.
 \eea

The diagonal part of the coefficients matrix in Eq.~(\ref{nmatrix3}) corresponds to a maximally entangled state, which yields vanishing energy density. Only the off-diagonal part can be responsible for  non-vanishing energies, and
its contribution scales as $N_{coh} C_{BH}$. This scaling follows from our previous treatments \cite{slowburn,flameoff},
where $N_{coh}C_{BH}$ has been identified as the effective (semiclassical) expansion parameter for perturbative calculations.
We then have $\;{\cal E}_{\rm pair} \simeq T_H^4 N_{coh} C_{BH}\;$.

The total energy density ${\cal E}$ is obtained by summing over all the coherent pairs, of which there are order $N_{coh}$,
\be
{\cal E} \simeq T_H^4 N_{coh}^2 C_{BH}\;.
\ee

In our model, $\;N_{coh}\sim S_{BH}^{1/2}\;$ \cite{slowburn}; however, in Hawking's model, $\;N_{coh}=0\;$ and, in the Page model,  $\;N_{coh}= S_{BH}\;$. To allow for all possibilities, let us parametrize $N_{coh}$ as follows:
\be
N_{coh}= (S_{BH})^{1-\alpha/2}\;.
\ee
So that, for Hawking's model, $\;\alpha\to \infty\;$, for the Page model, $\;\alpha= 0\;$ and, for our model, $\;\alpha=1\;$. Other choices of $\alpha$ interpolate between these different models. For $\;\alpha<2\;$, then $\;N_{coh}<1\;$,
and so these  models should behave in a similar way to Hawking's models for which $N_{coh}$ vanishes.

Now let us consider
\be
N_{coh}^2 C_{BH}=(S_{BH})^{1-\alpha}\;,
\ee
from which it follows that
\be
{\cal E}=  (S_{BH})^{1-\alpha}\ T_H^4\;.
\ee

The equivalence principle dictates that the in-falling shell sees a state similar to the vacuum. The deviations from the vacuum must be controlled by the size of the region being probed; in our case, $R_S$. (See \cite{bousso2} for a discussion in the current context.)

The near-horizon energy density ${\cal E}$ is clearly sensitive to the value of $\alpha$. For models with $\;0\le \alpha<1\;$, the energy density is parametrically larger than the Hawking temperature and, therefore, in contradiction with the dictate of the equivalence principle.  The energy density grows with the number of emitted particles $N_T$, scaling as $N_T^2 C_{BH}$. After the emission of $\sqrt{S_{BH}}$ particles, the energy density begins to differ significantly from $T_H^4$. Whether or not such regions of higher energy density  should be ascribed a firewall is probably a subjective matter. So far, no  quantitative  definition has been proffered (however, see \cite{bousso2}). We will suggest that such ambiguous cases be deemed as ``walls of smoke''. But, then again, where there's smoke, there's fire.

For models with $ \;\alpha\ge 1\;$, the equivalence principle seems to be satisfied. However, those
with $\;\alpha > 1\;$ effectively require the introduction of a new sub-Planckian gravitational scale that  can be expected to violate the equivalence principle. To understand why, we recall from \cite{slowburn} that $\;dR_S/dN_T=-l_p/\sqrt{S_{BH}}\;$, where $l_p$ is the Planck length and $N_T$ now means the number of emitted particles.
The difference in the horizon scale after the emission of $N_{coh}$ particles is given by $\;R_S(N_T+N_{coh})-R_S(N_T)\simeq \frac{dR_S}{dN_T} N_{coh}
=-(l_p/\sqrt{S_{BH}})N_{coh}\;$. In our model, this length scale is identified with the quantum width of the wavefunction,
$\;\sigma=l_p\;$. But, for general $\alpha$,
$\;\sigma=l_p \left(\frac{l_p}{R_S}\right)^{\alpha-1}$. Then $\;\sigma<l_p\;$ for $\;\alpha>1\;$.

Unitarity, on the other hand, can be preserved for a different range of $\alpha$. To see this, let us
 consider the information $I(N_T)$ that is released from the BH after the emission of  $N_T$ particles. Information is defined in the standard way as the difference between the thermal entropy and the actual entropy of the emitted radiation. We  recall from \cite{slowburn} that the information released from the BH is given by
\be
I(N_T)\sim N_T N_{coh} C_{BH}\;.
\ee

When $N_{coh} C_{BH}$ becomes order one, the rate of  release is reaching order unity and information starts to emerge rapidly from the BH. By this time, the transparency time, the amount of information released is of order  $S_{BH}(0)$. The earliest that this happens is for the Page model, where the amount released is $S_{BH}(0)/2$. For our model,
the transparency time occurs later but  the amount of information released by then is  larger,
$\;I\sim S_{BH}(0)-(S_{BH}(0))^{2/3}\sim S_{BH}(0)\;$. For arbitrary $\alpha$, using the estimates of \cite{slowburn}, we find  that  $\;dI/dN_T\simeq 1\;$ when $\;(S_{BH}(0)-N_T)^{2-\alpha/2}=(S_{BH}(0))^{2-\alpha}\;$.  And so
$\;S_{BH}(0)-N_T=(S_{BH}(0))^{2\frac{2-\alpha}{4-\alpha}}\;$. If $\;\alpha>2\;$, then $dI/dN_T$ never reaches 1 before the BH has completely evaporated, which is a consequence of $N_{coh}$ being less than 1.
 The conclusion is that such models do not provide the opportunity for all the  information to escape from the BH.

One can then see that Hawking's model satisfies the equivalence principle in an extreme way but unitarity is grossly violated. Page's model rather satisfies unitarity in an extreme manner, but it does not take into account the existence of a (semi)classical horizon, thus  sacrificing the equivalence principle. Our model preserves both
principles in what seems to be a unique way.

Let us now consider the perspective of an external observer and see how the equivalence principle constrains the rate of information emission in the various models. Here, we are considering a shell of matter of total mass $m$ such that $\;m\ll M_{BH}\;$. The in-falling shell is therefore  ``more quantum" than the BH, as its ratio of Compton wavelength to size $(\hbar/m)/R_{shell}$ is much larger than that of the BH when the shell is about to cross the horizon
({\em i.e.}, $\;R_{shell}\sim R_S$).

We want to show that the absorption time of the shell is the scrambling time $\sim R_S \ln R_S\;$. Although lacking a formal proof, the scrambling time can be viewed as a consequence of causality. Information on the horizon
cannot spread faster than unity ({\em i.e.}, the speed of light), and so the causal lower bound
on the time  for information to dissipate on the horizon is, locally, $R_S$. This bound can be translated to
that of an external observer's perspective by
using the standard relation between Kruskal and Schwarzschild  coordinates,  and one
readily obtains $R_S\ln{R_S}$.

The same  reasoning can be used to determine the absorption time of the shell. Here, there is another scale
to consider besides the size of the horizon, the quantum width of the wavefunction of the shell $\;\sigma\sim \hbar/m\ll R_S\;$. Given the same consideration of causality, the absorption time
of the shell as measured by an external observer is $\;R_S\ln{\sigma}\sim R_S\ln{\frac{\hbar}{m}}\;$, which is
shorter than the scrambling time by the logarithmic order $\;\ln{\frac{\hbar/m}{R_S}}\;$. The conclusion is that the absorption of a large (but quantum) shell cannot significantly alter the findings of HP regarding small
quantum objects. Consequently, if a  BH is to act as a perfect  reflector of such an object, what is necessary is that $k+c$ Hawking particles are capable of carrying $k+c$ bits of information \cite{HaydenPreskill}.

We can, however, see that this will  not be the generic outcome because the information release rate,
\be
\frac{dI}{dN_T}= N_{coh} C_{BH}\;,
\ee
is controlled by the same parameter $N_{coh} C_{BH}$ that controls the semiclassical approximation. And so $\;{dI}/{dN_T} \simeq 1\;$ occurs simultaneously with the breakdown of the semiclassical approximation.

The amount of information
released after some ``time''  $\dn$ is given by,
\be
\Delta I \;= \; \frac{dI}{dN_T} \dn
=
N_{coh} C_{BH}\; \dn\;.
\ee
So that, for the $\alpha$-models,
\be
\Delta I = \Delta N (S_{BH})^{-\alpha/2}\;.
\label{dndI}
\ee

As HP adopted the Page model for which $\;\alpha=0\;$, they indeed found that, after the Page time, the BH becomes an ``information mirror". Eq.~(\ref{dndI}) does demonstrate
just how sensitive HP's conclusion is to  this choice of model. For any other value of $\alpha$ differenent from zero, the
reflecting capability of the BH is suppressed.

\subsection*{Conclusion}

We have argued that the models of BH evaporation have to be evaluated according to their compatibility with two fundamental tenets, unitarity and the equivalence principle. These represent the two theories, quantum mechanics and general relativity, that provide the framework for discussing semiclassical gravitational physics.  We have found that a  range of models can comply with each of the principles separately; however, a unique model that is compliant with both  is singled out. This is our previously proposed model in which  $l_p$ and $R_S$ are the only scales. The Page model --- which is (at least implicitly) adopted universally in discussions about BH evaporation in general and  firewalls in particular --- is found to be incompatible with the equivalence principle. This is our interpretation of the firewall problem.

We have also discussed the differences between the various models by considering the plight of an in-falling shell of matter.  The external observer sees the shell thermalize;  after which, the information on its state  is carried out in the Hawking radiation. The rate at which information about the state of the shell emerges depends on the specific model,  but the chain of events is still the same; thermalization followed by radiation. On the other hand, what the shell sees depends dramatically on the model: The ordinary vacuum, a firewall or, perhaps, a  wall of smoke.

From the perspective of an external observer, the different models determine the ability of a BH to reflect quantum  matter. However, what is still true in any model compatible with unitarity is that  the BH can act as a perfect information mirror in the sense of Hayden and Preskill. This happens after the transparency time when information streams  rapidly out of the BH.

\section*{Acknowledgments}

We thank Ofer Aharony, Sam Braunstein and Sunny Itzhaki for useful discussions. The research of RB was supported by the Israel Science Foundation grant no. 239/10.
The research of AJMM received support from an NRF Incentive Funding Grant 85353,
an NRF KIC Grant 83407 and a Rhodes Discretionary Grant RD31/2013.
AJMM thanks Ben Gurion University
for their hospitality during his visits.

\end{document}